\documentclass{FAB-2018}  
\usepackage{hyperref}
\usepackage{url}
\usepackage{graphicx}
\usepackage{graphics} 
\usepackage{epsfig} 
\usepackage{mathptmx} 
 \usepackage{mathrsfs}
\usepackage{times} 
\usepackage{amsmath} 
\usepackage{amssymb}  
\usepackage{color}
\usepackage{float}
\usepackage{algorithm}
\usepackage{algorithmic}
\usepackage{etoolbox}

\usepackage{graphics} 
\usepackage{epsfig} 
\usepackage{mathptmx} 
 \usepackage{mathrsfs}
\usepackage{times} 
\usepackage{amsmath} 
\usepackage{amssymb}  
\usepackage{color}
\usepackage{float}
\usepackage{algorithm}
\usepackage{algorithmic}
\usepackage{hyperref}
\usepackage{url}
\usepackage{calrsfs}
\DeclareMathAlphabet{\pazocal}{OMS}{zplm}{m}{n}

\makeatletter
\def\hlinew#1{%
  \noalign{\ifnum0=`}\fi\hrule \@height #1 \futurelet
   \reserved@a\@xhline}
\patchcmd{\@makecaption}
{\scshape}
{}
{}
{}
\makeatother 

\begin{document}

	\title{Towards Trusted Social Networks with Blockchain Technology}

	\numberofauthors{3} 
	%
	\author{
		\alignauthor
		Yize Chen\\
		\affaddr{University of Washington}\\
		\affaddr{Seattle, WA, USA}\\
		\email{yizechen@uw.edu}
		\alignauthor
		Quanlai Li\\
		\affaddr{University of California, Berkeley}\\
		\affaddr{Berkeley, CA, USA}\\
		\email{quanlai\_li@berkeley.edu}
		\alignauthor
		Hao Wang\\
		\affaddr{University of Washington}\\
		\affaddr{Seattle, WA, USA}\\
		\email{hwang16@uw.edu}
	}

	\date{15 Dec 2017}

	\maketitle
	\begin{abstract}
		Large-scale rumor spreading could pose severe social and economic damages. The emergence of online social networks along with the new media can even make rumor spreading more severe. Effective control of rumor spreading is of theoretical and practical significance. This paper takes the first step to understand how the blockchain technology can help limit the spread of rumors. 
		Specifically, we develop a new paradigm for social networks embedded with the blockchain technology, which employs decentralized contracts to motivate trust networks as well as secure information exchange contract. We design a blockchain-based sequential algorithm which utilizes virtual information credits for each peer-to-peer information exchange. We validate the effectiveness of the blockchain-enabled social network on limiting the rumor spreading. Simulation results validate our algorithm design in avoiding rapid and intense rumor spreading, and motivate better mechanism design for trusted social networks.
	
	\end{abstract}
	
	
	
	
	\section{Introduction}
	\label{intro}
	Rumor has been existing for thousands of years in human history. A rumor often refers to a piece of unverified information~(e.g., explanation of events, media coverage, and information exchange) circulating from person to person or pertaining to an object, event, or issue of public concern~\cite{peterson1951rumor}.  In the age of the Internet, denser connections among individuals along with faster information transmission rate also trigger rapid rumor propagation, and could cause more intense social panics and negative effects~\cite{bordia1996studying}. 

Past studies have put emphasis on both the modeling techniques and the avoidance mechanisms of rumor spreading. Yet considering the complexities of rumor transmission dynamics, the diversity of social networks, along with the emergence of information transmission media, most of these studies cannot find the root of rumor blast nor a general yet effective approach to eliminate rumor dissemination~\cite{doerr2012rumors}.  


The blockchain technology then becomes a good fit, which has seen its success in financial area for trusted and secure contracts. This has motivated us to re-design the information exchange process as a ``contract"-based process in modern social networks. In addition, the pair-wise spreading style of rumor also lets blockchain-based contract to become a good fit for future information propagation and exchange platform.

In this work, we introduce a mechanism for smart contract design that makes full use of the expressive power fulfilled by blockchain technologies. By allocating a virtual accumulated credit for each member in the social network, we design an innovative approach for information exchange. Such credits are a reflection of the credibility of both social network members and corresponding information. The proposed algorithm is designed to avoid the spreading of ``untrusted information", which is information without sufficient endorsement.

\begin{figure}[h]
    \centering
    \includegraphics[scale=0.35]{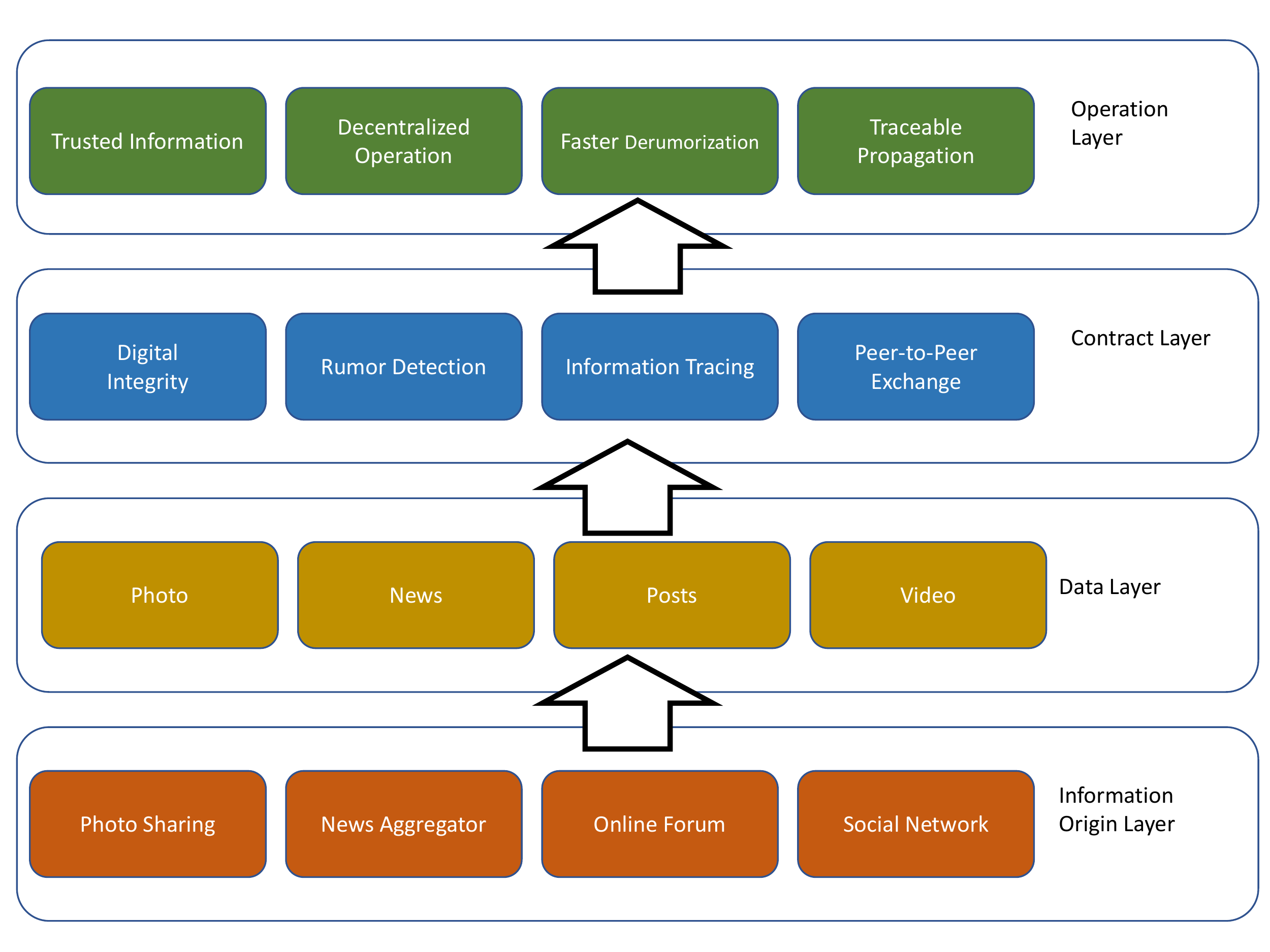}
    \caption{\small
        {
            The architecture of blockchain-enabled information exchange system.
        }
    }
    \label{architecture}
\end{figure}

To illustrate that such a mechanism design would help to avoid the large-scale propagation of fake news through the network, we design and set up a graphical model along with the nonlinear systems for the social networks that are of interests. We show that for peer-to-peer information exchange and propagation, individuals under blockchain are more cautious about the authenticity of the information. Our simulation examined the propagation of information with and without the proposed mechanism, and showed that our proposed approach can effectively reduce the social and economic damage by rumor. To our knowledge, this paper is the first work aiming to utilize the characteristics of blockchain to address and solve the rumor spreading problems in social networks. The general architecture for our proposed approach is shown in Fig.~\ref{architecture}.

\subsection{Related Work}

\subsubsection{Rumor Spreading Model}
A rumor is a piece of \emph{unverified circulating information}. Past research on rumor involved multidisciplinary efforts from physics, sociology, and psychology. Several approaches to the modeling of rumor spreading and control of its damage were discussed~\cite{allport1947psychology, rosnow1976rumor}. Previous rumor models regarded the heterogeneous social network as a graph where rumors propagate. Studies in~\cite{moreno2004dynamics} used stochastic processes to simulate rumor spreading to get a better understanding. In~\cite{vega2017rumor}, the authors observed that more influential spreaders exist on social networks. They assigned higher probabilities for them to spread the information. In the context of new types of social media and networks (e.g., micro-blogging), studies in~\cite{zhao2013sir} proposed a SIR~(Susceptible, Infected, Recovered) rumor spreading model. In this model, the spreading process is classified as susceptible, infected, and recovered. The work is based on the assumption that ignorants are easily influenced by the spreader, and that accordance with reality will change the probabilities of converting a spreader into a stifler. 

Previous studies have shown that the cessation or blast of rumors is mainly related to the stifling and forgetting mechanisms for a given network~\cite{daley1964epidemics, nekovee2007theory}. New forms of social networks, such as bidirectional information exchanges, also emerge. In this case, the receiver could also have an influence on the spreader. We will leverage the blockchain technology in this type of social network, and examine how this will affect the spread of rumors~(e.g., change of immunity).

\subsubsection{Blockchain Technology}
A blockchain is a linked chain of growing list of blocks~\cite{brito2013bitcoin}. Every block contains its corresponding record and the timestamp. The blockchain is designed with a peer-to-peer network, where each node propagates its records to other nodes. This design prevents unvalidated modification of data. 

Researchers have implemented blockchain-based protocols to build a decentralized network~\cite{zyskind2015decentralizing}. In the network, the third party is replaced by an automated access-control manager, enabled by the distributed blockchain system. Other researchers proposed to adopt blockchain in supply chain management for a better quality~\cite{chen2017blockchain}. Blockchain can solve the traceability and trustability problems in this scenario. People also find blockchain useful in power grid industry~\cite{basden2017utilities}. Both utilities and consumers benefit from this technology by recording and validating the information on a distributed network affordably and reliably. Meanwhile, a combination of blockchain and the internet of things (IoT) increases utilization of cloud storage~\cite{shafagh2017towards}. The blockchain is also suitable for other applications, such as online transaction, identity management, notarization~\cite{zyskind2015decentralizing}. 

The main contributions of this paper are as follows:
\begin{itemize}
    \item We propose an innovative decentralized mechanism for social network information exchange based on blockchain;
    \item We build a blockchain-enabled SIR model and show from numerical simulations that from the ``regulator" perspective proposed algorithm could effectively control rumor spreading on social networks.
\end{itemize}

The remainder of the paper is organized as follows. In Section~2, we present the rumor spreading model for social networks. In Section~3, we develop the blockchain-enabled architecture for social networks. In Section~4, we conduct numerical simulations to validate the effectiveness of our blockchain-enabled algorithm, and Section~5 concludes the paper.

	\section{Rumor Spreading Model}
	\label{model}
	In this section, we present the SIR epidemiological model, which can be used to characterize the rumor spreading dynamics for a social network with a group of fixed participants. We analyze the temporal characteristics of such a stochastic model~(e.g., the peak value, the convergence rate and the final state), and introduce the potential roles that the blockchain technology could play to reduce the spread of rumors. We also discuss several practical issues in real-world social networks.

\begin{figure*}[h]
    \centering
    \includegraphics[scale=1.08]{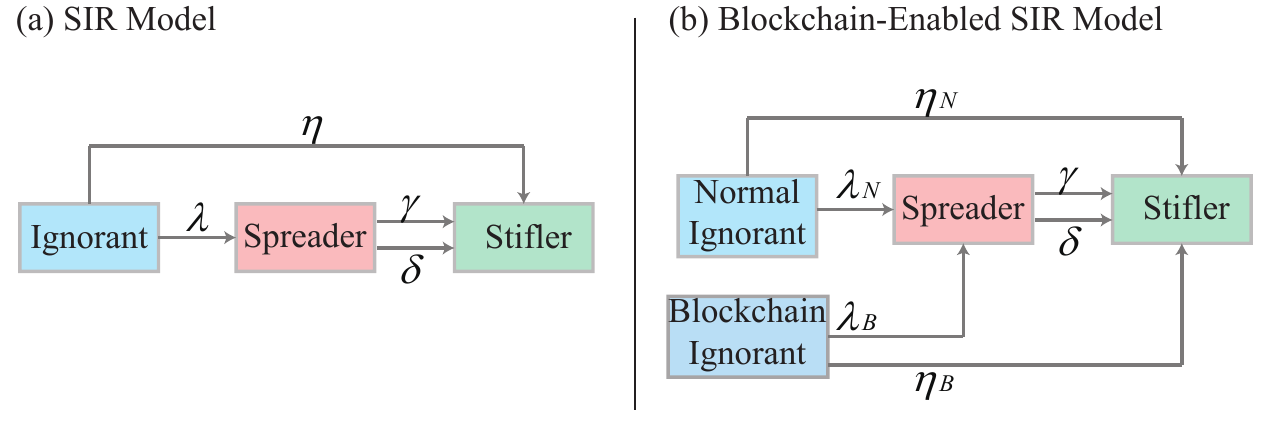}
    \caption{\small
        {
            The comparison of classic SIR model for social networks~(Fig. \ref{schematic}a) and SIR model under blockchain technology~(Fig. \ref{schematic}b). A new group, the blockchain-enabled ignorants $I_B$, are emerged and taking part in the information exchange.
        }
    }
    \label{schematic}
\end{figure*}

\subsection{Model Setup}

Consider an undirected graph $G=(V, E)$, where $V$ is a set of vertices representing individuals in the social network, and $E$ is a set of edges representing the social interactions. We assume that the social network has a fixed number of homogeneously mixed population, and the degree distribution for nodes in $G$ conforms to the Poisson distribution:
\begin{equation}
P(k)=e^{-\bar{k}} \frac{\bar{k}^k}{k!},
\end{equation}
where $\bar{k}$ is the average degree for $G$ and $P(k)$ denotes the probability of observing $k$ degrees for $v \in V$.

To better investigate how rumors are propagated through the network, we adopt a rule-based classification method that divides the vertices into three convertible sets~\cite{zhao2013sir, nekovee2007theory, beretta1995global}: the spreader set $S$, the ignorant set $I$, and the stifler set $R$. The dynamics of these three classes are as follows:
\begin{itemize}
    \item \emph{Ignorant with Density $I(t)$}. The ignorants are similar to susceptible individuals in classic SIR models. At time $t>0$, an ignorant has a probability $\lambda$ to become a spreader when it has contact with a spreader who is quite certain of the truth of the rumor. Afterwards, it's willing to spread the rumor in the following time steps. Meanwhile, the ignorant has probability $\eta$ to become a stifler, who has no interests in the rumor anymore.
    \item \emph{Spreader with Density $S(t)$}. A spreader is contributing to the propagation of rumor within $G$. Any spreader involved in a pair-wise meeting attempts to ``infect" other individuals with the known rumor. At time $t>0$, when a spreader contacts with a stifler, the spreader has a probability $\gamma$ to convert to a stifler. In addition, we take the forgetting mechanism~\cite{nekovee2007theory} into consideration and assume that at a certain time, a spreader itself has forgotten the rumor and then turns into a stifler at rate $\delta$.
    \item \emph{Stifler with Density $R(t)$}. A stifler is contributing to the final elimination of the rumor. In general, it is an absorbing state in our stochastic model, and are accumulating its density by turning both ignorants and spreaders into stiflers.
\end{itemize}

To summarize all the dynamics considered above, we derive a nonlinear system consisting of the following differential equations for $I(t), S(t)$ and $R(t)$, respectively.

\begin{subequations}
    \begin{align}
    \label{model1}
    \frac{dI(t)}{dt}&=-(\lambda + \eta) \bar{k}I(t)S(t),\\
\frac{dS(t)}{dt}&=\lambda \bar{k} I(t)S(t) - \gamma \bar{k}S(t)(S(t)+R(t))-\delta S(t),\\
\frac{dR(t)}{dt}&=\eta \bar{k} I(t)S(t)+\gamma \bar{k}S(t)(S(t)+R(t))+\delta S(t),
    \end{align}
\end{subequations}
with the corresponding model structure plotted in Fig.~\ref{schematic}a.

\subsection{System Dynamics and Practical Issues}
We initialize a social network $G$ with $|V|=N$ with $a$ spreaders who know the rumor and are willing to spread:
\begin{equation}
I(0)=\frac{N-a}{N}, \quad S(0)=\frac{a}{N}, \quad R(0)=0.
\end{equation}

We are interested in the dynamics of the rumor spreading model, e.g., the peak density of spreaders and the velocity of rumor spreading. It is also shown in~\cite{zhao2013sir} that when the system approaches to the final states, there are only ignorants and stiflers left in the network, while spreaders for untrusted information will die out. It is then important to observe the final state of $R(t)$, since a smaller $R(t)$ indicates that when the rumor appears again, the group of $I(t)$ will have to face the rumor spreading issue throughout the network.

The applicability of the model described in Section.~\ref{model1} is also justified in several previous studies~\cite{nekovee2007theory, jin2013epidemiological}. In \cite{jin2013epidemiological} it showed that such model is well fitted for real Twitter data on a set of real-world news (e.g., Boston Marathon Bombings and Pope Resignation).

	\section{System Architecture}
	\label{system}
	In this section, we first describe the proposed blockchain-enabled protocol for information exchange, which can be integrated into the social network model described in Section.~\ref{model}. We then illustrate how such a blockchain-enabled algorithm can propel a trusted social network as a whole.

\subsection{Blockchain Protocols for Rumor Spreading}

To ensure both security and privacy of the information change process and avoid large-scale spreading of untrusted piece of messages, we adopt the blockchain technology and design a protocol consisting of \emph{private contract} and \emph{public contract}. We allocate an accumulated virtual information credit for each participant in the network, and use such credits to motivate the propagation of trusted information.

\subsubsection{Private contract} The private contract is negotiated and signed between the spreader and the receiver offline. Consider a group of spreaders $s_i\in S$ and a receiver $r$. At timestep $t$, validation between $s_i$ and $r$ is executed before a private contract is negotiated. For example, once the receiver $r$'s desires have been accomplished, it ``pays" virtual credit $cred_{rs}(t)$ to the spreader $s_i$~(denoted as $cred_{s_i}$), while the spreader is in charge of sending the information $info_{sr}(t)$ to the receiver $r$~(denoted as $info_r$). This ``investment" of credit can pay off once this piece of information is validated to be trustworthy. The accumulated credits increase for receiver $r$. On the contrary, once the information is validated as a rumor, the accumulated credits of receiver $r$ would decrease. 

In Algorithm \ref{private_contract}, we illustrate the working principle of such a peer-to-peer information-credit exchange program.

 \begin{algorithm}
 \label{private}
    \caption{Private Smart Contract}
    \label{private_contract}
    \begin{algorithmic}
        \ENSURE {Initial spreader set $S$}
        \ENSURE{$info \leftarrow \emptyset$, $cred \leftarrow \emptyset$}
        \REQUIRE {Contract Receiver $r$, $r$'s accumulated credit $c(t)$}
        \FOR {$s_i$ in connection of $r$}
            \IF {$s_i \in S$} 
                \STATE{\# \emph{$s_i$ and $r$ form a secure channel to negotiate contract}}
                \IF {Contract made}
                    \STATE{$c(t+1)=c(t)-cred_{rs}(t)$}
                    \STATE{$info_r \leftarrow info_{s_i r}(t)$, $cred_{s_i} = cred_{s_i} + cred_{r s_i}(t)$}
                    \PRINT
                \ENDIF
            \ENDIF
        \ENDFOR
    \end{algorithmic}
\end{algorithm}    

\subsubsection{Public contract}
The public contract is updated at every time step to record the links of information propagation as well as the credit flows throughout the social network. It serves as the public ledger for all information transactions. A transaction on information is used as evidence of contractor consent. This contract also makes the highest transaction credit $C_{max}$ public to all existing participants of the information exchange, which is available for decision-making in private contract negotiation stage. The logs recording each transaction time, credits along with the hash form the block. The pseudocode for achieving such a chain of contract is sketched in Algorithm \ref{public_contract}.

 \begin{algorithm}
    \caption{Public Smart Contract}
    \label{public_contract}
    \begin{algorithmic}
        \ENSURE {highest credit $C_{max}=0$; Credit of each member $cred_i$}
        \ENSURE {$info_{list} \leftarrow \emptyset$, $cred_{list} \leftarrow \emptyset$}
        \FOR {$t\in T$}
        \IF {Contract made}
        \STATE {Update $C_{max}$ through the network}
        \STATE{Update $cred_{list}$, $info_{list}$}
        \ENDIF
        \ENDFOR
    \end{algorithmic}
\end{algorithm}

By employing the two-layer contract design for information exchange, we are able to construct a distributed, synchronized contract network which is secure and resilient. Moreover, as the network evolves, $C_{max}$ increases with respect to the network consensus, which indicates either higher risk or higher credibility for ignorant to trust given information. Such public transaction information would guide each member under blockchain contract make their private decisions. Moreover, note that our blockchain-based contract architecture is not closed only for contractors. For normal individuals in the social networks, they possess their original information exchange process. In Section.~\ref{simulation} we compare the simulation results with different ratio of blockchain-based individuals involved in the social networks.

\subsection{Rumor Spreading Model Under Blockchain}
We begin by firstly introducing a trusted model along with the modeling assumptions. Then we justify how the blockchain technology would help inhibit the propagation of rumors on social networks.

The only difference between the model described here and the model described in Section.~\ref{model} is that we have a group of initial social network participants who have signed a blockchain-enabled trust contract. For the simplicity, we denote the density of initial members of such contract as $I_B$ and initial members without signing the contract as $I_N$. Note that $I_N$ conforms to the similar ignorants' dynamics as described in \ref{model1}, and the corresponding probability of converting to a spreader or a stifler is $\lambda_N$ and $\eta_N$, respectively. Whenever there is information exchange between two individuals with at least one individual belonging to $I_B$, they run the secure and reliable consensus protocol to agree upon the pre-defined virtual credits for members under the trust contract. Since individual coming from $I_B$ has access to the public information provided by all existing blockchain contracts, she/he has a different estimate of virtual credits of information exchange. Therefore, $I_B$ have different dynamics compared to $I_N$, and we denote the corresponding probability as $\lambda_B$ and $\eta_B$, respectively. We can derive the dynamics of $I_N(t)$, $I_B(t)$, $S(t)$ and $R(t)$ on the blockchain-enabled social networks:

\begin{subequations}
    \begin{align}
    \label{model2}
    \frac{dI_B(t)}{dt}&=-(\lambda_B+\eta_B)\bar{k}I_B(t)S(t),\\
    \frac{dI_N(t)}{dt}&=-(\lambda_N+\eta_N)\bar{k}I_N(t)S(t), \\
    \frac{dS(t)}{dt}&=\lambda_B \bar{k} I_B(t)S(t)+\lambda_N \bar{k} I_N(t)S(t) - \gamma \bar{k}S(t)(S(t)+R(t))-\delta S(t),\\
    \frac{dR(t)}{dt}&=\eta_B \bar{k} I_B(t)S(t)+\eta_N \bar{k} I_N(t)S(t)+\gamma \bar{k}S(t)(S(t)+R(t))+\delta S(t).
    \end{align}
\end{subequations}

\subsection{Mechanisms Analysis and Discussion}
\label{discussion}
\emph{Rumor Spreading Rate}: Once an individual has signed the trust contract under blockchain protocols, it has an extra record coming from the blockchain ``transactions" list of information within the whole network. This gives her/him an additional estimate of the ``value" for possible information exchange measured by accumulated credits, and thus can make a better judgment of the authenticity of information based on the risk of losing credits in certain transactions.

In general, once the private contract between two individuals has reached a higher value of the virtual credit, members from $I_B$ are more cautious about the ongoing information exchange. Then members from $I_B$ are less vulnerable when exposing to a rumor. Meanwhile, they are more likely to lose interest in a rumor and thus convert to stiflers directly.

Hence, with blockchain-enabled contract signed, ignorants from $I_B$ have limited contributions to the spreader $S$ but more contributions to the stifler $R$:
\begin{equation}
\lambda_B<\lambda_N, \quad \eta_B > \eta_N.
\end{equation}

\begin{figure}[h]
    \centering
    \includegraphics[scale=1.0]{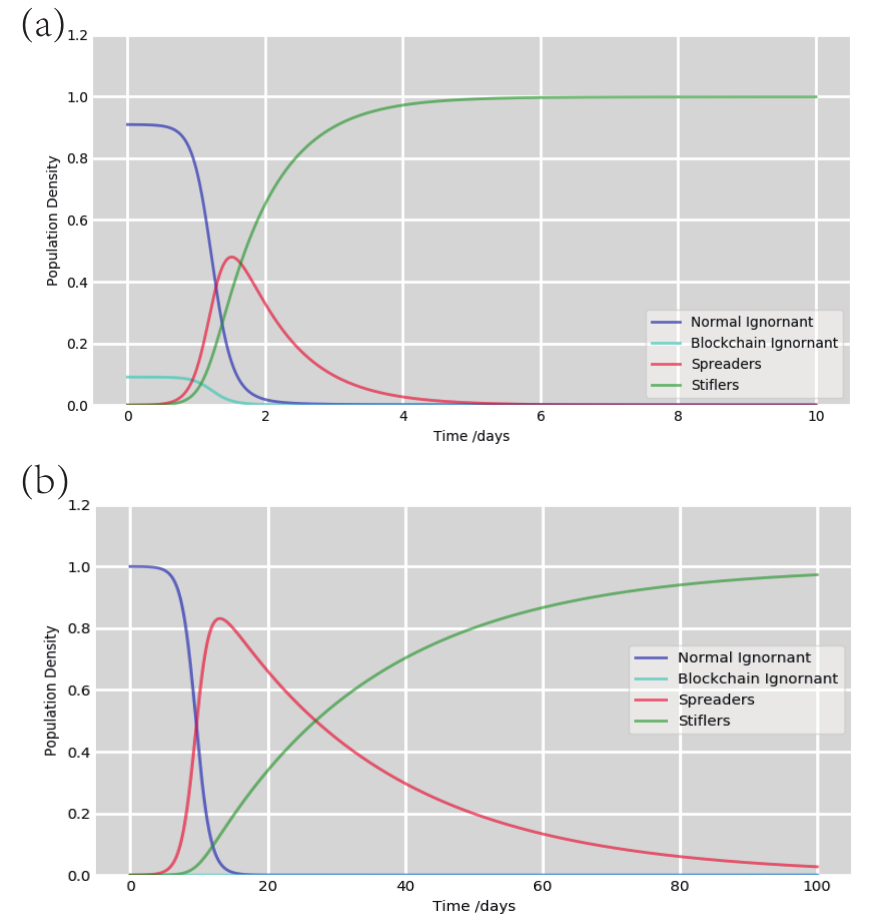}
    \caption{\small
        {
            The temporal dynamics of the proposed blockchain-enabled system (Fig.~\ref{timeseries}a) compared to a blockchain-free system~(Fig.~\ref{timeseries}b). Blockchain ignorants are under the blockchain contracts throughout the time.
        }
    }
    \label{timeseries}
\end{figure}

\begin{figure*}[h]
    \centering
    \includegraphics[scale=0.45]{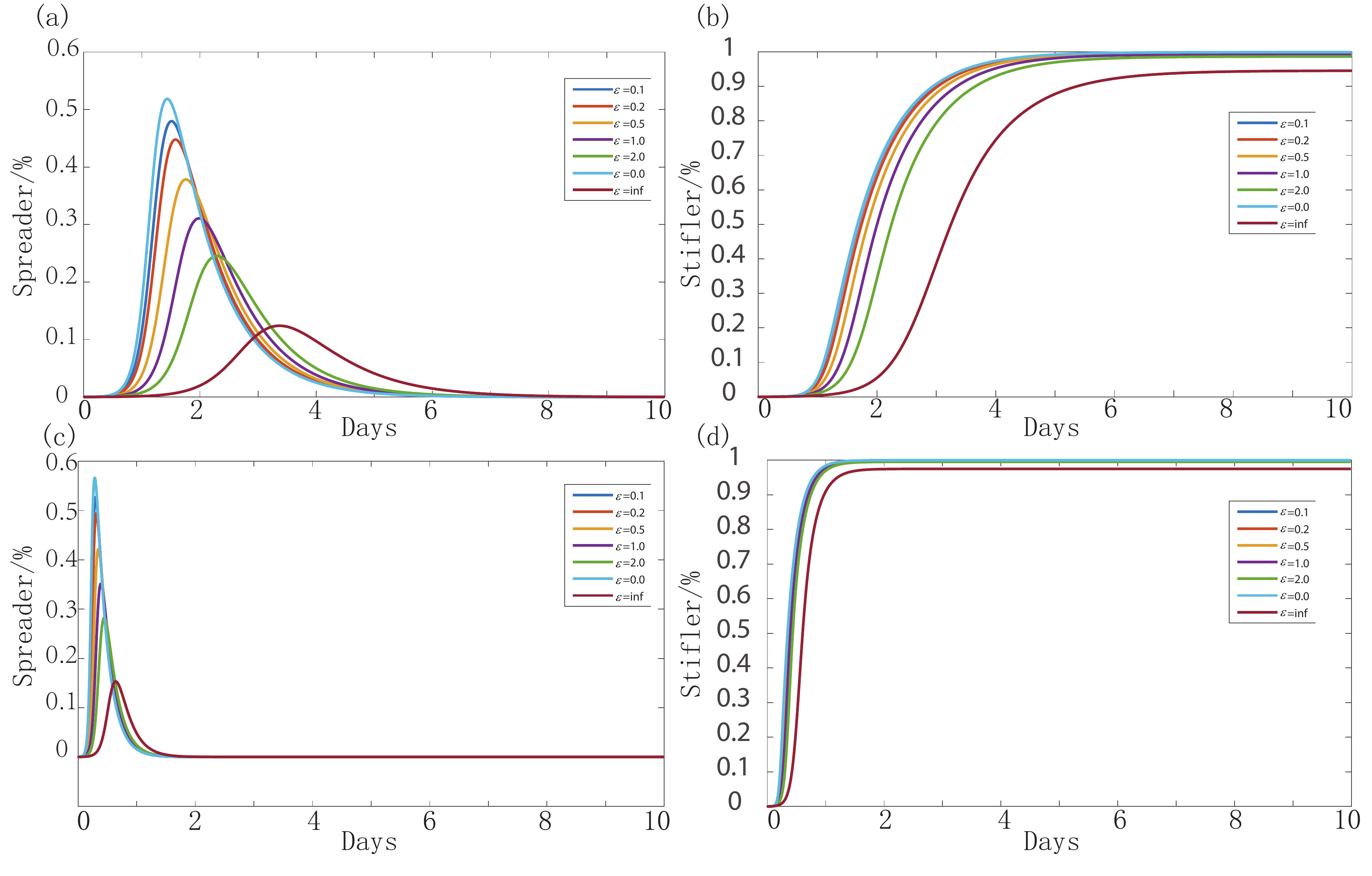}
    \caption{\small
        {
            The temporal dynamics of the spreader density $\frac{S(t)}{N}$ (Fig.~\ref{temporal}a and Fig.~\ref{temporal}c) and the stifler density $\frac{R(t)}{N}$ (Fig.~\ref{temporal}b and Fig.~\ref{temporal}d) under different $\epsilon$. 
        }
    }
    \label{temporal}
\end{figure*}

\emph{Forgetting Mechanism}: with blockchain-enabled SIR model for social networks, the forgetting mechanism not only takes account of the ``forget" process of spreaders, it also takes account of those spreaders under blockchain contract, who are less likely to keep a ``fake" news in their public records. So with blockchain technology enabled in a given $G$, $\delta$ tends to approach a higher value than the model in \ref{model1}. This indicates that the social network could get a higher absorbing rate from spreaders to stiflers.

	\section{Numerical Simulations}
	\label{simulation}
	Following the model introduced in Section.\ref{model}, in this section, we conduct numerical simulations to validate the performance of blockchain-enabled rumor spreading on social network models, and compare the results with blockchain-free rumor spreading performance.

We are particularly interested in the rumor spreading process from the initial stage $t=0$ till the terminal stage $t=T$. That is to say, in all our simulations over $t \in \{0,1,...,T\}$, we start with $S(0)=1, \, R(0)=0$ out of a fixed overall population of $S(t)+R(t)+I(t)=10000$, where $I(t)=I_B(t)+I_N(t)$. Note that we constrain our simulation to the case that stifler is the final absorbing state. Based on the discussion in Section.~\ref{discussion}, we set $\lambda_B=0.3,\, \lambda_N=0.8, \, \eta_B=0.7, \, \eta_N=0.2$ to investigate the influence of blockchain contract. To better evaluate the performance of the blockchain contract in our model, we introduce $\epsilon=\frac{I_B(0)}{I_N(0)}$ to control the initial population ratio. We also consider two group of settings for $\bar{k}$, where  $\bar{k}=10$ corresponds to a sparse, traditional information exchange platform, and $\bar{k}=50$ corresponds to a dense, newly-emerged information exchange platform.

In the first experiment, we investigate two situations for the dynamics of rumor spreading process, which are shown in Fig.~\ref{timeseries}. In Fig.~\ref{timeseries}a we show a group of setting with $\epsilon=0.1, \, \delta=0.3, \, \gamma = 0.1, \, \bar{k}=10$. In Fig.~\ref{timeseries}b, we simulate the extreme circumstance in which no member signs the blockchain contract with $\epsilon=0$, while members from $I_N$ easily trust the rumors with $\lambda_N=0.99,\, \eta_N=0.01, \, \bar{k}=10$. We observe from Fig.~\ref{timeseries}b that the rumor exists much longer on the social networks~(over 100 days compared with less than 6 days in Fig.~\ref{timeseries}a). Moreover, the peak density for spreaders is over $81\%$, which indicates that most of the members easily trust rumors. In contrast, with a portion of the population enrolled in the blockchain contract, the peak value of rumor density can be cut down to $48\%$ as is shown in Fig.~\ref{timeseries}a.

\begin{figure}[h]
    \centering
    \includegraphics[scale=1.1]{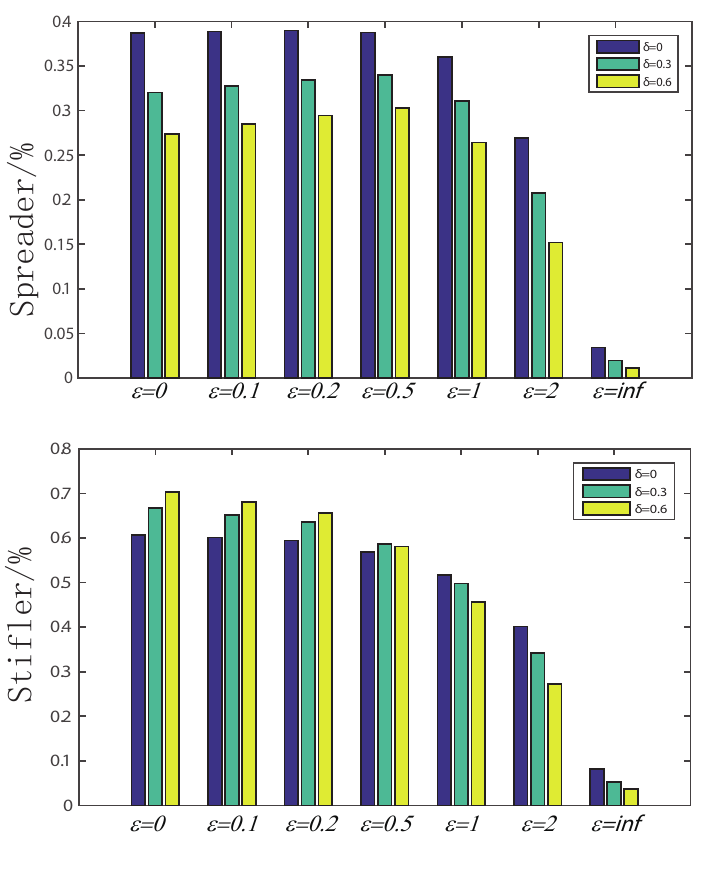}
    \caption{\small
        {
            The spreader density $\frac{S(t)}{N}$ (Fig.~\ref{bar}a) and the stifler density $\frac{R(t)}{N}$ (Fig.~\ref{bar}b) at the end of day $2$ under different $\epsilon$ and $\delta$. \vspace{-15pt}
        }
    }
    \label{bar}
\end{figure}

We then conduct simulations to evaluate the impact of the number of individuals who sign the blockchain contract, as depicted in Fig.~\ref{temporal}. Fig.~\ref{temporal}a and Fig.~\ref{temporal}b show the simulation results under $\bar{k}=10$, implying everyone on the social network possesses relatively sparse connections~(e.g., information exchange through newspapers and phone calls). Fig.~\ref{temporal}c and Fig.~\ref{temporal}d are simulated under $\bar{k}=50$, in which everyone on the social network is densely connected~(e.g., information exchange through online social networks). Note that $\epsilon=\inf$ indicates the scenario where all the individuals except the initial spreader have signed the blockchain contract. 

From Fig.~\ref{temporal}a and Fig.~\ref{temporal}c, we observe that as the number of initial blockchain contractors increases (larger $\epsilon$ value), the peak value of spreader density drops significantly. In addition, the peak is deferred compared to the case with a smaller $\epsilon$ value, indicating that the rumor spreading process has been delayed. This delayed and weaken rumor spreading process also provides an opportunity for external intervention (e.g., a credible clarification) to control rumor spreading. One interesting finding is that if we consider a dense social network which is more prevalent in today's new media as well as online social networks, the rumor spreading process is much quicker~(Fig.~\ref{temporal}c) compared to traditional rumor spreading media. Results depicted in Fig.~\ref{temporal}b and Fig.~\ref{temporal}d also verify that with a higher penetration of blockchain-enabled members, initial ignorants are more skeptical to the rumors and thus take a longer time to finally convert to stiflers. In addition, the results also show that $\frac{R(T)}{N}<1$ decreases as $\epsilon$ increases in the terminal state.

Our final simulation evaluates the mixed impact of $\epsilon$ and $\delta$. Since the blockchain contract would also trigger some converted spreaders to be skeptical of rumors, $\delta$ can increase, and spreaders ``forget" rumors and finally convert to stiflers. In Fig.~\ref{bar}a and Fig.~\ref{bar}b, we show the density of spreaders and stiflers at the end of day 2, respectively. We observe that a larger $\delta$ would drag down the spreader density, because a larger $\delta$ represents the spreaders are more willing to reconsider the rumor as fake information and convert to stiflers. Meanwhile, a larger $\epsilon$ can eventually drag down the spreader density, but as we observed in Fig.~\ref{temporal}a and Fig.~\ref{temporal}c, a larger $\epsilon$ would also change the rumor spreading speed. Therefore, there is no significant change in spreader density when $\epsilon$ is relatively small. 

Results shown in Fig.~\ref{bar} also motivate the mechanism design for a blockchain-based information propagation contract. The design of an appropriate virtual credit would not only control the system dynamics~(e.g., rumor spreading speed and the peak value), but also regulate each participant's behavior~(e.g., distributing different initial information credits).  
	
	\section{Conclusion and Discussion}
	In this work, we investigated the dynamics of rumor dissemination in social networks with and without blockchain-enabled technology. We firstly introduced the graphical model setup for social networks. We then illustrated how to incorporate the blockchain contract into peer-to-peer information exchange process by employing virtual credits. The re-designed blockchain-enabled rumor spreading model along with numerical simulation demonstrated that blockchain technology would help in avoiding large-scale rumor spreading. Such model setup and simulation results would motivate us to design trust-based information exchange system with blockchain technology enabled.

In the future work, we would also like to inspect the extreme case that is not included in this work, e.g., at initial point, most of members are spreaders, or during the information propagation, members are with low immunity. Contracts designed for extreme conditions and large-scale social networks may be designed and considered.
	
\bibliographystyle{IEEEtran}
\bibliography{bib}

\end{document}